%% file: gsh-inel.tex
\documentstyle[twoside,fleqn,espcrc2]{article}

\newcommand{\AmS}{{\protect\the\textfont2
  A\kern-.1667em\lower.5ex\hbox{M}\kern-.125emS}}

\title{Inelasticity for hadron-carbon nucleus collisions from 
       emulsion chamber}

\author{G. Wilk\address{The Andrzej Soltan Institute for Nuclear Studies, 
        Nuclear Theory Department, Warsaw, Poland\\
        email: wilk@fuw.edu.pl}
        and 
        Z.W\l odarczyk\address{Institute of Physics, Pedagogical 
        University, Kielce, Poland \\
        email: wlod@pu.kielce.pl}%
	\thanks{Talk given  at the Xth International Symposium
	on Very High Energy Cosmic Ray Interactions, Laboratori
	Nazionali del Gran Sasso, 12-17 July 1998, Assergi,
	Italy, to be published in the proceedings ({\sl Nucl.
	Phys. {\bf B} (Proc. Suppl.)).}}}
       
\begin{document}

\begin{abstract}
The inelasticity of hadron-carbon collisions for energies exceeding
$100$ TeV is estimated from the carbon-emulsion chamber data at
Pamirs to be $\langle K_C\rangle = 0.65\pm 0.08$. When combined with
data on hadron-lead collisions taken at the same energy range it
results in the $K\sim A^{0.086}$ mass number dependence of
inelasticity. The evaluated partial inelasticity for secondary ($\nu
> 1$) interactions, $K_{\nu >1} \simeq 0.2$, suggests that most of
the energy is lost in the first interaction.
\end{abstract}

\maketitle

\section{INTRODUCTION}

The inelasticity of hadronic reactions, understood as the fraction of
the incident beam energy spent on the production of secondaries, is
(next to the inelastic cross section) the most significant variable
for all cosmic ray experiments involved in cascade developments
\cite{SWWW,FGS}. Unfortunately, in the energy region exceeding a few
hundreds of GeV there are no accelerator data on inelasticity on
nuclear targets and only rough indications from cosmic ray
experiments are available \cite{SWWW,FGS}. Recently \cite{B} the
inelasticity in hadron-lead collisions was estimated in the energy
region exceeding $100$ TeV. In this contribution we present similar
analysis performed for $h-C$ collisions observed in carbon emulsion
chamber (EC) exposed to cosmic rays at the Pamirs. 

This EC consists of $\Gamma$-block of $6$ cm $Pb$ ($0.35 \lambda$,
$10.5$ c.u.) and two $H$-blocks of carbon layer of $60$ cm thickness
($66 {\rm g}/{\rm cm}^2, 0.9\lambda, 2.5$ c.u.) each followed by $5$
cm of lead-emulsion sandwiches. In EC (which is a shallow
calorimeter) only the energy transfered to the electromagnetic
component is measured, i.e., $E_h ^{\gamma}\, =\, K_{\gamma} \cdot
E_h$, and in the hadronic block a given nuclear-electro\-ma\-gnetic
cascade (NEC) produces spots with optical density $D$ on $X$-ray film
(cf., \cite{TW} for details).

Such structure of the carbon EC allows for a relatively
straightforward estimation of the total inelasticity for $h-C$
interactions \cite{NWP}. The proposed method is uses the repeated
registration of the same cascade in the two subsequent hadronic
blocks. If $N_1$ denote the number of cascades registered in   
the first ha\-dro\-nic block with visible energy $E_1 > (E_h^{\gamma})_1$
and $N_2$ the number of cascades repeated registered in second
hadronic block with $E_2$ above the threshold ($E_h^{\gamma})_2$, then
the ratios 

\begin{equation}
\eta \, =\, \frac{N_2}{N_1}\qquad {\rm and}\qquad 
\epsilon\, =\, \frac{E_2}{E_1}
\end{equation}

\noindent
are sensitive to total inelasticity $K$. The weak dependence of these
quantities on the methodical errors and ease with which the
experimental data may by obtained render this method very useful and
promising for possible future applications.  

\section{INELASTICITY FOR CARBON TARGET}

The experimental data collected from $110 {\rm m}^2$ carbon EC
contain $N_1 = 70$ hadrons with energies $E_1 > 30$ TeV and
$N_2 = 24$ hadrons with energies $E_2 > 2$ TeV). They give the value
of $\eta = 0.27\pm 0.06$ (at energy threshold $E_2 > 4$ TeV, being free
from the detection bias) and the energy ratio $\epsilon = 0.24\pm
0.07$. These data have been then recalculated by using the simulated
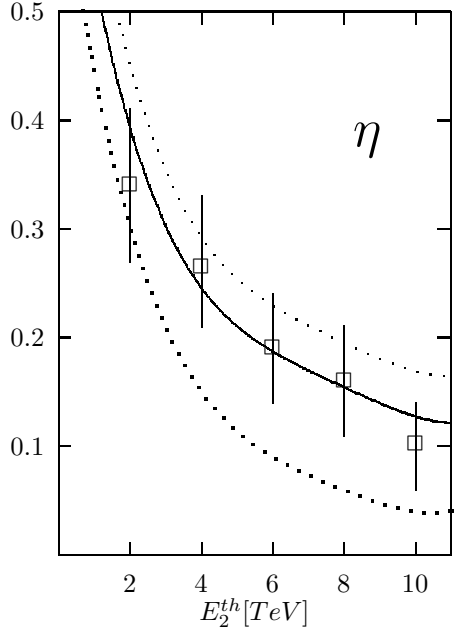
\begin{figure}[htb]
\vspace{9pt}
\input{fig1_inel}
\caption{
Dependence of $\eta = N_2/N_1$ on the energy threshold
$E{th}_2$ in the second hadronic block for: $\langle K\rangle = 0.65$
(solid line), $\langle K\rangle =0.50$ (dotted line) and 
for $\langle K\rangle = 0.80$ (black dots), compared with the 
experimental data.
}
\label{fig:1}
\end{figure}
$D(E_h^{\gamma})$ dependence \cite{TW}. The repeated registrations of
hadron has been simulated by the Monte-Carlo event generator. Primary
hadrons (assumed to consist of $75\%$ nucleons and $25\%$ pions) were
sampled from the power spectrum representing distribution of the
initial energy with a differential slope equal to $\gamma = 3$. The
gamma quanta and electrons above $0.01$ TeV, reaching the detection
level within the radius of $5$ mm, were recorded and the
corresponding optical densities were calculated within the radii
utilized in the experiment. Only cascades with the energies above
$E_1 = 30$ TeV and $E_2 = 2$ TeV were selected.

The ratio $\eta$ of the number of hadrons repeatedly registered in
two hadronic blocks and the number of all hadrons registered in the
first hadro\-nic block is presented in Fig.1 for different total
inelasticities: $\langle K\rangle = 0.5, 0.65$ and $0.80$. Note that
the ratio $\eta$ is more sensitive to the mean value of inelasticity
$\langle K\rangle$ than the energy ratio $\epsilon$, shown in Fig.2.
The comparison of experimental data with simulated dependences
indicates that $\langle K_C\rangle = 0.65 \pm 0.08$ for hadron-carbon
nucleus collisions at the hadron energies of above $\sim 100$ TeV is
the most probably choice for the mean value of inelasticity for
hadron-carbon collisions.
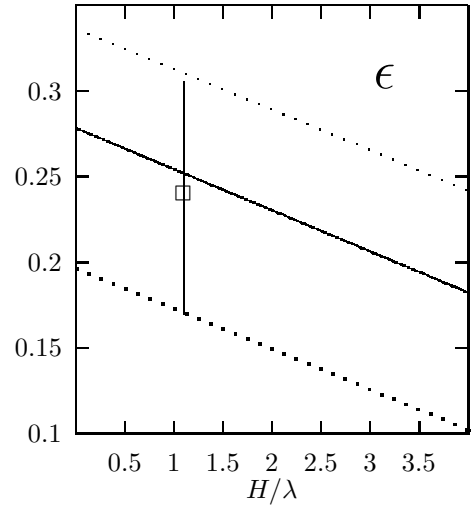
\begin{figure}[htb]
\input{fig2_inel}
\caption{
Dependence of $\epsilon = E_2/E_1$ on the thickness
$H/\lambda$ of carbon target (the plotted curves correspond to
different $\langle K\rangle$ as in Fig. 1).
}
\label{fig:2}
\end{figure}

This result can be now compared with $\langle K_{Pb}\rangle = 0.83
\pm 0.17$ obtained in similar analysis of succesive hadron
interactions registered in the thick-lead-emulsion chamber at Pamirs
\cite{B}. Both results lead to the $K \simeq A^{0.086}$ mass number
dependence of inelasticity. If we extrapolate this $A$ dependence to
$A=1$ we obtain hadron-proton inelasticity in this energy range being equal
to $K_{h-p} = 0.53$ (notice that the same value can be obtained from
Eq. (\ref{eq:K2}) below where $K_1 = K_{h-p}$). Assuming further that
$N_{\pi}/N_h \sim 0.25$ and that $K_{\pi-p} = 1.5 K_{p-p}$, we obtain
that in the energy range exceeding $100$ TeV inelasticity in
proton-proton collisions is equal to $K_{p-p} = 0.46$ (which agrees
with our earlier predictions \cite{SWWW,Ohsawa2}).

\section{PARTIAL INELASTICITY $K_{\nu}$}

Following Ref. \cite{FGS} we shall now estimate the so called partial
inelasticity $K_{\nu}$. In the framework of Glauber multiple
scattering formalism \cite{G}) it is defined by

\begin{equation}
\langle 1 - K\rangle \, =\, \sum_{\nu = 1}\,  P_{\nu} \langle 1 -
K_{\nu}\rangle  ^{\nu}
\end{equation}

\noindent
where $P_{\nu}$ is the probability for encountering exactly $\nu$
wounded nucleons in a target of mass $A$ and $\langle 1 - K_{\nu}
\rangle$  is the mean elasticity of the leading hadron in collisions
with exactly $\nu$ wounded nucleons. Assuming that $K_{\nu >1} = K_2$
the total elasticity can be written as

\begin{equation}
\langle 1 - K\rangle \, =\, \left( 1 - K_1\right)\, \sum_{\nu = 1}
\langle 1 - K_2 \rangle ^{\nu - 1}\,  P_{\nu } .\label{eq:K2}
\end{equation}

\noindent
The ratio $\kappa$ of elasticities (given by eq.(\ref{eq:K2})) in
collisions on Pb and C targets depends only on $K_2$ once the
$P_{\nu}$ is known.  Assuming Poisson distribution for the number of
wounded nucleons, $\nu $, the value of $K_2$ for the expected mean
number of wounded nucleons $\langle \nu \rangle  = \frac{A
\sigma_{h-p}}{\sigma_{hA}} \sim A^{1/3}$  and for the experimatally
evaluated value of $\kappa = 0.5$ is equal to $K_2 \simeq 0.2$.
(Notice tacid assumption made here that the ultimate identity of the
final state nucleon is determined only once during the interaction
with the nucleus - in \cite{FGS} it means that $\beta = 1$ for the
parameter specifing the fraction of isospin preserving reactions).

\section{CONCLUDING REMARKS}

For hadron-carbon nucleus collisions in energy region exceeding
$100$ TeV the inelasticity is estimated to be equal to $\langle
K_C\rangle = 0.65\pm 0.08$. Our estimation of $K_2 = 0.2$ at energies
above $100$ TeV is consistent with low energy data (cf. Ref.
\cite{FGS}). Note that inequality $K_{\nu > 1} < K_1$ is
characteristic to all string-type interaction models (cf. Quark-Gluon
String model \cite{QGS} or Dual Parton Model \cite{DPM}) whereas the
SIBYLL model \cite{FGS,SIBYLL} predict much smaller value of $K_2$ in
the examined energy region. It is important to notice that the new
results for inelasticity on lead target, $\langle K_{h-Pb}\rangle
\simeq 0.6 \pm 0.05$ reported at this conference \cite{Ohsawa} when
compared with our $\langle K_{h-C}\rangle$ result in $K_2 \simeq 0$
and $\langle K_{h-p}\rangle = \langle K_{h-C}\rangle$ which exceeds
noticeably the value $\langle K_{pp}\rangle \simeq 0.41$ as evaluated
from the collider data \cite{Ohsawa2}.

\end{document}

%% file: fig1_inel.tex
\setlength{\unitlength}{0.240900pt}
\ifx\plotpoint\undefined\newsavebox{\plotpoint}\fi
\sbox{\plotpoint}{\rule[-0.200pt]{0.400pt}{0.400pt}}%
\begin{picture}(900,990)(0,0)
\font\gnuplot=cmr10 at 10pt
\gnuplot
\sbox{\plotpoint}{\rule[-0.200pt]{0.400pt}{0.400pt}}%
\put(220.0,113.0){\rule[-0.200pt]{148.394pt}{0.400pt}}
\put(220.0,113.0){\rule[-0.200pt]{0.400pt}{205.729pt}}
\put(220.0,284.0){\rule[-0.200pt]{4.818pt}{0.400pt}}
\put(198,284){\makebox(0,0)[r]{0.1}}
\put(816.0,284.0){\rule[-0.200pt]{4.818pt}{0.400pt}}
\put(220.0,455.0){\rule[-0.200pt]{4.818pt}{0.400pt}}
\put(198,455){\makebox(0,0)[r]{0.2}}
\put(816.0,455.0){\rule[-0.200pt]{4.818pt}{0.400pt}}
\put(220.0,625.0){\rule[-0.200pt]{4.818pt}{0.400pt}}
\put(198,625){\makebox(0,0)[r]{0.3}}
\put(816.0,625.0){\rule[-0.200pt]{4.818pt}{0.400pt}}
\put(220.0,796.0){\rule[-0.200pt]{4.818pt}{0.400pt}}
\put(198,796){\makebox(0,0)[r]{0.4}}
\put(816.0,796.0){\rule[-0.200pt]{4.818pt}{0.400pt}}
\put(220.0,967.0){\rule[-0.200pt]{4.818pt}{0.400pt}}
\put(198,967){\makebox(0,0)[r]{0.5}}
\put(816.0,967.0){\rule[-0.200pt]{4.818pt}{0.400pt}}
\put(332.0,113.0){\rule[-0.200pt]{0.400pt}{4.818pt}}
\put(332,68){\makebox(0,0){2}}
\put(332.0,947.0){\rule[-0.200pt]{0.400pt}{4.818pt}}
\put(444.0,113.0){\rule[-0.200pt]{0.400pt}{4.818pt}}
\put(444,68){\makebox(0,0){4}}
\put(444.0,947.0){\rule[-0.200pt]{0.400pt}{4.818pt}}
\put(556.0,113.0){\rule[-0.200pt]{0.400pt}{4.818pt}}
\put(556,68){\makebox(0,0){6}}
\put(556.0,947.0){\rule[-0.200pt]{0.400pt}{4.818pt}}
\put(668.0,113.0){\rule[-0.200pt]{0.400pt}{4.818pt}}
\put(668,68){\makebox(0,0){8}}
\put(668.0,947.0){\rule[-0.200pt]{0.400pt}{4.818pt}}
\put(780.0,113.0){\rule[-0.200pt]{0.400pt}{4.818pt}}
\put(780,68){\makebox(0,0){10}}
\put(780.0,947.0){\rule[-0.200pt]{0.400pt}{4.818pt}}
\put(220.0,113.0){\rule[-0.200pt]{148.394pt}{0.400pt}}
\put(836.0,113.0){\rule[-0.200pt]{0.400pt}{205.729pt}}
\put(220.0,967.0){\rule[-0.200pt]{148.394pt}{0.400pt}}
\put(705,765){\makebox(0,0){\huge{$\eta$}}}
\put(528,23){\makebox(0,0){$E^{th}_2 [TeV]$}}
\put(220.0,113.0){\rule[-0.200pt]{0.400pt}{205.729pt}}
\put(311.00,967.00){\usebox{\plotpoint}}
\put(315.32,946.73){\usebox{\plotpoint}}
\multiput(320,930)(4.844,-20.182){2}{\usebox{\plotpoint}}
\put(330.84,886.44){\usebox{\plotpoint}}
\put(336.25,866.40){\usebox{\plotpoint}}
\put(341.88,846.42){\usebox{\plotpoint}}
\put(348.31,826.69){\usebox{\plotpoint}}
\put(354.80,806.98){\usebox{\plotpoint}}
\put(361.05,787.19){\usebox{\plotpoint}}
\put(367.75,767.55){\usebox{\plotpoint}}
\put(375.81,748.43){\usebox{\plotpoint}}
\multiput(376,748)(7.288,-19.434){0}{\usebox{\plotpoint}}
\put(383.19,729.03){\usebox{\plotpoint}}
\put(391.07,709.83){\usebox{\plotpoint}}
\put(399.25,690.76){\usebox{\plotpoint}}
\multiput(400,689)(9.840,-18.275){0}{\usebox{\plotpoint}}
\put(408.82,672.35){\usebox{\plotpoint}}
\put(418.47,653.98){\usebox{\plotpoint}}
\multiput(419,653)(9.939,-18.221){0}{\usebox{\plotpoint}}
\put(428.82,636.00){\usebox{\plotpoint}}
\multiput(432,631)(11.513,-17.270){0}{\usebox{\plotpoint}}
\put(440.07,618.56){\usebox{\plotpoint}}
\multiput(444,612)(11.513,-17.270){0}{\usebox{\plotpoint}}
\put(451.37,601.17){\usebox{\plotpoint}}
\multiput(456,595)(13.668,-15.620){0}{\usebox{\plotpoint}}
\put(464.45,585.07){\usebox{\plotpoint}}
\multiput(469,579)(13.508,-15.759){0}{\usebox{\plotpoint}}
\put(477.57,569.00){\usebox{\plotpoint}}
\multiput(481,565)(15.759,-13.508){0}{\usebox{\plotpoint}}
\put(492.43,554.57){\usebox{\plotpoint}}
\multiput(494,553)(14.676,-14.676){0}{\usebox{\plotpoint}}
\multiput(500,547)(14.676,-14.676){0}{\usebox{\plotpoint}}
\put(507.21,540.00){\usebox{\plotpoint}}
\multiput(512,536)(16.889,-12.064){0}{\usebox{\plotpoint}}
\put(523.54,527.22){\usebox{\plotpoint}}
\multiput(525,526)(15.945,-13.287){0}{\usebox{\plotpoint}}
\multiput(531,521)(17.270,-11.513){0}{\usebox{\plotpoint}}
\put(540.12,514.77){\usebox{\plotpoint}}
\multiput(544,512)(17.270,-11.513){0}{\usebox{\plotpoint}}
\multiput(550,508)(17.270,-11.513){0}{\usebox{\plotpoint}}
\put(557.30,503.13){\usebox{\plotpoint}}
\multiput(562,500)(18.564,-9.282){0}{\usebox{\plotpoint}}
\multiput(568,497)(18.021,-10.298){0}{\usebox{\plotpoint}}
\put(575.28,492.81){\usebox{\plotpoint}}
\multiput(581,489)(18.564,-9.282){0}{\usebox{\plotpoint}}
\multiput(587,486)(18.564,-9.282){0}{\usebox{\plotpoint}}
\put(593.41,482.77){\usebox{\plotpoint}}
\multiput(600,479)(18.564,-9.282){0}{\usebox{\plotpoint}}
\put(611.77,473.11){\usebox{\plotpoint}}
\multiput(612,473)(18.564,-9.282){0}{\usebox{\plotpoint}}
\multiput(618,470)(18.564,-9.282){0}{\usebox{\plotpoint}}
\put(630.51,464.21){\usebox{\plotpoint}}
\multiput(631,464)(18.564,-9.282){0}{\usebox{\plotpoint}}
\multiput(637,461)(18.564,-9.282){0}{\usebox{\plotpoint}}
\multiput(643,458)(18.564,-9.282){0}{\usebox{\plotpoint}}
\put(649.09,454.96){\usebox{\plotpoint}}
\multiput(656,452)(18.564,-9.282){0}{\usebox{\plotpoint}}
\put(667.84,446.08){\usebox{\plotpoint}}
\multiput(668,446)(18.564,-9.282){0}{\usebox{\plotpoint}}
\multiput(674,443)(18.564,-9.282){0}{\usebox{\plotpoint}}
\put(686.58,437.18){\usebox{\plotpoint}}
\multiput(687,437)(19.690,-6.563){0}{\usebox{\plotpoint}}
\multiput(693,435)(18.564,-9.282){0}{\usebox{\plotpoint}}
\multiput(699,432)(18.564,-9.282){0}{\usebox{\plotpoint}}
\put(705.54,428.85){\usebox{\plotpoint}}
\multiput(712,427)(18.564,-9.282){0}{\usebox{\plotpoint}}
\multiput(718,424)(18.564,-9.282){0}{\usebox{\plotpoint}}
\put(724.59,420.80){\usebox{\plotpoint}}
\multiput(730,419)(18.564,-9.282){0}{\usebox{\plotpoint}}
\multiput(736,416)(19.957,-5.702){0}{\usebox{\plotpoint}}
\put(744.01,413.66){\usebox{\plotpoint}}
\multiput(749,412)(18.564,-9.282){0}{\usebox{\plotpoint}}
\multiput(755,409)(19.690,-6.563){0}{\usebox{\plotpoint}}
\put(763.37,406.32){\usebox{\plotpoint}}
\multiput(768,405)(19.690,-6.563){0}{\usebox{\plotpoint}}
\multiput(774,403)(20.473,-3.412){0}{\usebox{\plotpoint}}
\put(783.35,400.88){\usebox{\plotpoint}}
\multiput(786,400)(20.473,-3.412){0}{\usebox{\plotpoint}}
\multiput(792,399)(19.957,-5.702){0}{\usebox{\plotpoint}}
\put(803.53,396.24){\usebox{\plotpoint}}
\multiput(805,396)(20.473,-3.412){0}{\usebox{\plotpoint}}
\multiput(811,395)(20.756,0.000){0}{\usebox{\plotpoint}}
\multiput(817,395)(20.547,-2.935){0}{\usebox{\plotpoint}}
\put(824.12,394.00){\usebox{\plotpoint}}
\multiput(830,394)(20.756,0.000){0}{\usebox{\plotpoint}}
\put(836,394){\usebox{\plotpoint}}
\multiput(288.59,957.71)(0.485,-2.171){11}{\rule{0.117pt}{1.757pt}}
\multiput(287.17,961.35)(7.000,-25.353){2}{\rule{0.400pt}{0.879pt}}
\multiput(295.59,927.84)(0.482,-2.480){9}{\rule{0.116pt}{1.967pt}}
\multiput(294.17,931.92)(6.000,-23.918){2}{\rule{0.400pt}{0.983pt}}
\multiput(301.59,900.39)(0.482,-2.299){9}{\rule{0.116pt}{1.833pt}}
\multiput(300.17,904.19)(6.000,-22.195){2}{\rule{0.400pt}{0.917pt}}
\multiput(307.59,874.67)(0.482,-2.208){9}{\rule{0.116pt}{1.767pt}}
\multiput(306.17,878.33)(6.000,-21.333){2}{\rule{0.400pt}{0.883pt}}
\multiput(313.59,850.89)(0.485,-1.789){11}{\rule{0.117pt}{1.471pt}}
\multiput(312.17,853.95)(7.000,-20.946){2}{\rule{0.400pt}{0.736pt}}
\multiput(320.59,826.22)(0.482,-2.027){9}{\rule{0.116pt}{1.633pt}}
\multiput(319.17,829.61)(6.000,-19.610){2}{\rule{0.400pt}{0.817pt}}
\multiput(326.59,803.50)(0.482,-1.937){9}{\rule{0.116pt}{1.567pt}}
\multiput(325.17,806.75)(6.000,-18.748){2}{\rule{0.400pt}{0.783pt}}
\multiput(332.59,781.77)(0.482,-1.847){9}{\rule{0.116pt}{1.500pt}}
\multiput(331.17,784.89)(6.000,-17.887){2}{\rule{0.400pt}{0.750pt}}
\multiput(338.59,761.05)(0.482,-1.756){9}{\rule{0.116pt}{1.433pt}}
\multiput(337.17,764.03)(6.000,-17.025){2}{\rule{0.400pt}{0.717pt}}
\multiput(344.59,742.08)(0.485,-1.408){11}{\rule{0.117pt}{1.186pt}}
\multiput(343.17,744.54)(7.000,-16.539){2}{\rule{0.400pt}{0.593pt}}
\multiput(351.59,722.60)(0.482,-1.575){9}{\rule{0.116pt}{1.300pt}}
\multiput(350.17,725.30)(6.000,-15.302){2}{\rule{0.400pt}{0.650pt}}
\multiput(357.59,704.88)(0.482,-1.485){9}{\rule{0.116pt}{1.233pt}}
\multiput(356.17,707.44)(6.000,-14.440){2}{\rule{0.400pt}{0.617pt}}
\multiput(363.59,688.16)(0.482,-1.395){9}{\rule{0.116pt}{1.167pt}}
\multiput(362.17,690.58)(6.000,-13.579){2}{\rule{0.400pt}{0.583pt}}
\multiput(369.59,672.79)(0.485,-1.179){11}{\rule{0.117pt}{1.014pt}}
\multiput(368.17,674.89)(7.000,-13.895){2}{\rule{0.400pt}{0.507pt}}
\multiput(376.59,656.43)(0.482,-1.304){9}{\rule{0.116pt}{1.100pt}}
\multiput(375.17,658.72)(6.000,-12.717){2}{\rule{0.400pt}{0.550pt}}
\multiput(382.59,641.71)(0.482,-1.214){9}{\rule{0.116pt}{1.033pt}}
\multiput(381.17,643.86)(6.000,-11.855){2}{\rule{0.400pt}{0.517pt}}
\multiput(388.59,627.99)(0.482,-1.123){9}{\rule{0.116pt}{0.967pt}}
\multiput(387.17,629.99)(6.000,-10.994){2}{\rule{0.400pt}{0.483pt}}
\multiput(394.59,614.99)(0.482,-1.123){9}{\rule{0.116pt}{0.967pt}}
\multiput(393.17,616.99)(6.000,-10.994){2}{\rule{0.400pt}{0.483pt}}
\multiput(400.59,602.74)(0.485,-0.874){11}{\rule{0.117pt}{0.786pt}}
\multiput(399.17,604.37)(7.000,-10.369){2}{\rule{0.400pt}{0.393pt}}
\multiput(407.59,590.54)(0.482,-0.943){9}{\rule{0.116pt}{0.833pt}}
\multiput(406.17,592.27)(6.000,-9.270){2}{\rule{0.400pt}{0.417pt}}
\multiput(413.59,579.54)(0.482,-0.943){9}{\rule{0.116pt}{0.833pt}}
\multiput(412.17,581.27)(6.000,-9.270){2}{\rule{0.400pt}{0.417pt}}
\multiput(419.59,568.82)(0.482,-0.852){9}{\rule{0.116pt}{0.767pt}}
\multiput(418.17,570.41)(6.000,-8.409){2}{\rule{0.400pt}{0.383pt}}
\multiput(425.59,559.21)(0.485,-0.721){11}{\rule{0.117pt}{0.671pt}}
\multiput(424.17,560.61)(7.000,-8.606){2}{\rule{0.400pt}{0.336pt}}
\multiput(432.59,549.09)(0.482,-0.762){9}{\rule{0.116pt}{0.700pt}}
\multiput(431.17,550.55)(6.000,-7.547){2}{\rule{0.400pt}{0.350pt}}
\multiput(438.59,540.09)(0.482,-0.762){9}{\rule{0.116pt}{0.700pt}}
\multiput(437.17,541.55)(6.000,-7.547){2}{\rule{0.400pt}{0.350pt}}
\multiput(444.59,531.37)(0.482,-0.671){9}{\rule{0.116pt}{0.633pt}}
\multiput(443.17,532.69)(6.000,-6.685){2}{\rule{0.400pt}{0.317pt}}
\multiput(450.59,523.37)(0.482,-0.671){9}{\rule{0.116pt}{0.633pt}}
\multiput(449.17,524.69)(6.000,-6.685){2}{\rule{0.400pt}{0.317pt}}
\multiput(456.59,515.69)(0.485,-0.569){11}{\rule{0.117pt}{0.557pt}}
\multiput(455.17,516.84)(7.000,-6.844){2}{\rule{0.400pt}{0.279pt}}
\multiput(463.59,507.65)(0.482,-0.581){9}{\rule{0.116pt}{0.567pt}}
\multiput(462.17,508.82)(6.000,-5.824){2}{\rule{0.400pt}{0.283pt}}
\multiput(469.00,501.93)(0.491,-0.482){9}{\rule{0.500pt}{0.116pt}}
\multiput(469.00,502.17)(4.962,-6.000){2}{\rule{0.250pt}{0.400pt}}
\multiput(475.59,494.65)(0.482,-0.581){9}{\rule{0.116pt}{0.567pt}}
\multiput(474.17,495.82)(6.000,-5.824){2}{\rule{0.400pt}{0.283pt}}
\multiput(481.00,488.93)(0.581,-0.482){9}{\rule{0.567pt}{0.116pt}}
\multiput(481.00,489.17)(5.824,-6.000){2}{\rule{0.283pt}{0.400pt}}
\multiput(488.00,482.93)(0.491,-0.482){9}{\rule{0.500pt}{0.116pt}}
\multiput(488.00,483.17)(4.962,-6.000){2}{\rule{0.250pt}{0.400pt}}
\multiput(494.00,476.93)(0.599,-0.477){7}{\rule{0.580pt}{0.115pt}}
\multiput(494.00,477.17)(4.796,-5.000){2}{\rule{0.290pt}{0.400pt}}
\multiput(500.00,471.93)(0.599,-0.477){7}{\rule{0.580pt}{0.115pt}}
\multiput(500.00,472.17)(4.796,-5.000){2}{\rule{0.290pt}{0.400pt}}
\multiput(506.00,466.93)(0.599,-0.477){7}{\rule{0.580pt}{0.115pt}}
\multiput(506.00,467.17)(4.796,-5.000){2}{\rule{0.290pt}{0.400pt}}
\multiput(512.00,461.93)(0.710,-0.477){7}{\rule{0.660pt}{0.115pt}}
\multiput(512.00,462.17)(5.630,-5.000){2}{\rule{0.330pt}{0.400pt}}
\multiput(519.00,456.93)(0.599,-0.477){7}{\rule{0.580pt}{0.115pt}}
\multiput(519.00,457.17)(4.796,-5.000){2}{\rule{0.290pt}{0.400pt}}
\multiput(525.00,451.94)(0.774,-0.468){5}{\rule{0.700pt}{0.113pt}}
\multiput(525.00,452.17)(4.547,-4.000){2}{\rule{0.350pt}{0.400pt}}
\multiput(531.00,447.94)(0.774,-0.468){5}{\rule{0.700pt}{0.113pt}}
\multiput(531.00,448.17)(4.547,-4.000){2}{\rule{0.350pt}{0.400pt}}
\multiput(537.00,443.94)(0.920,-0.468){5}{\rule{0.800pt}{0.113pt}}
\multiput(537.00,444.17)(5.340,-4.000){2}{\rule{0.400pt}{0.400pt}}
\multiput(544.00,439.94)(0.774,-0.468){5}{\rule{0.700pt}{0.113pt}}
\multiput(544.00,440.17)(4.547,-4.000){2}{\rule{0.350pt}{0.400pt}}
\multiput(550.00,435.94)(0.774,-0.468){5}{\rule{0.700pt}{0.113pt}}
\multiput(550.00,436.17)(4.547,-4.000){2}{\rule{0.350pt}{0.400pt}}
\multiput(556.00,431.94)(0.774,-0.468){5}{\rule{0.700pt}{0.113pt}}
\multiput(556.00,432.17)(4.547,-4.000){2}{\rule{0.350pt}{0.400pt}}
\multiput(562.00,427.95)(1.132,-0.447){3}{\rule{0.900pt}{0.108pt}}
\multiput(562.00,428.17)(4.132,-3.000){2}{\rule{0.450pt}{0.400pt}}
\multiput(568.00,424.94)(0.920,-0.468){5}{\rule{0.800pt}{0.113pt}}
\multiput(568.00,425.17)(5.340,-4.000){2}{\rule{0.400pt}{0.400pt}}
\multiput(575.00,420.95)(1.132,-0.447){3}{\rule{0.900pt}{0.108pt}}
\multiput(575.00,421.17)(4.132,-3.000){2}{\rule{0.450pt}{0.400pt}}
\multiput(581.00,417.95)(1.132,-0.447){3}{\rule{0.900pt}{0.108pt}}
\multiput(581.00,418.17)(4.132,-3.000){2}{\rule{0.450pt}{0.400pt}}
\multiput(587.00,414.95)(1.132,-0.447){3}{\rule{0.900pt}{0.108pt}}
\multiput(587.00,415.17)(4.132,-3.000){2}{\rule{0.450pt}{0.400pt}}
\multiput(593.00,411.94)(0.920,-0.468){5}{\rule{0.800pt}{0.113pt}}
\multiput(593.00,412.17)(5.340,-4.000){2}{\rule{0.400pt}{0.400pt}}
\multiput(600.00,407.95)(1.132,-0.447){3}{\rule{0.900pt}{0.108pt}}
\multiput(600.00,408.17)(4.132,-3.000){2}{\rule{0.450pt}{0.400pt}}
\multiput(606.00,404.95)(1.132,-0.447){3}{\rule{0.900pt}{0.108pt}}
\multiput(606.00,405.17)(4.132,-3.000){2}{\rule{0.450pt}{0.400pt}}
\multiput(612.00,401.95)(1.132,-0.447){3}{\rule{0.900pt}{0.108pt}}
\multiput(612.00,402.17)(4.132,-3.000){2}{\rule{0.450pt}{0.400pt}}
\multiput(618.00,398.95)(1.132,-0.447){3}{\rule{0.900pt}{0.108pt}}
\multiput(618.00,399.17)(4.132,-3.000){2}{\rule{0.450pt}{0.400pt}}
\multiput(624.00,395.95)(1.355,-0.447){3}{\rule{1.033pt}{0.108pt}}
\multiput(624.00,396.17)(4.855,-3.000){2}{\rule{0.517pt}{0.400pt}}
\multiput(631.00,392.95)(1.132,-0.447){3}{\rule{0.900pt}{0.108pt}}
\multiput(631.00,393.17)(4.132,-3.000){2}{\rule{0.450pt}{0.400pt}}
\multiput(637.00,389.95)(1.132,-0.447){3}{\rule{0.900pt}{0.108pt}}
\multiput(637.00,390.17)(4.132,-3.000){2}{\rule{0.450pt}{0.400pt}}
\multiput(643.00,386.95)(1.132,-0.447){3}{\rule{0.900pt}{0.108pt}}
\multiput(643.00,387.17)(4.132,-3.000){2}{\rule{0.450pt}{0.400pt}}
\put(649,383.17){\rule{1.500pt}{0.400pt}}
\multiput(649.00,384.17)(3.887,-2.000){2}{\rule{0.750pt}{0.400pt}}
\multiput(656.00,381.95)(1.132,-0.447){3}{\rule{0.900pt}{0.108pt}}
\multiput(656.00,382.17)(4.132,-3.000){2}{\rule{0.450pt}{0.400pt}}
\multiput(662.00,378.95)(1.132,-0.447){3}{\rule{0.900pt}{0.108pt}}
\multiput(662.00,379.17)(4.132,-3.000){2}{\rule{0.450pt}{0.400pt}}
\multiput(668.00,375.95)(1.132,-0.447){3}{\rule{0.900pt}{0.108pt}}
\multiput(668.00,376.17)(4.132,-3.000){2}{\rule{0.450pt}{0.400pt}}
\multiput(674.00,372.95)(1.132,-0.447){3}{\rule{0.900pt}{0.108pt}}
\multiput(674.00,373.17)(4.132,-3.000){2}{\rule{0.450pt}{0.400pt}}
\multiput(680.00,369.95)(1.355,-0.447){3}{\rule{1.033pt}{0.108pt}}
\multiput(680.00,370.17)(4.855,-3.000){2}{\rule{0.517pt}{0.400pt}}
\put(687,366.17){\rule{1.300pt}{0.400pt}}
\multiput(687.00,367.17)(3.302,-2.000){2}{\rule{0.650pt}{0.400pt}}
\multiput(693.00,364.95)(1.132,-0.447){3}{\rule{0.900pt}{0.108pt}}
\multiput(693.00,365.17)(4.132,-3.000){2}{\rule{0.450pt}{0.400pt}}
\multiput(699.00,361.95)(1.132,-0.447){3}{\rule{0.900pt}{0.108pt}}
\multiput(699.00,362.17)(4.132,-3.000){2}{\rule{0.450pt}{0.400pt}}
\multiput(705.00,358.95)(1.355,-0.447){3}{\rule{1.033pt}{0.108pt}}
\multiput(705.00,359.17)(4.855,-3.000){2}{\rule{0.517pt}{0.400pt}}
\put(712,355.17){\rule{1.300pt}{0.400pt}}
\multiput(712.00,356.17)(3.302,-2.000){2}{\rule{0.650pt}{0.400pt}}
\multiput(718.00,353.95)(1.132,-0.447){3}{\rule{0.900pt}{0.108pt}}
\multiput(718.00,354.17)(4.132,-3.000){2}{\rule{0.450pt}{0.400pt}}
\multiput(724.00,350.95)(1.132,-0.447){3}{\rule{0.900pt}{0.108pt}}
\multiput(724.00,351.17)(4.132,-3.000){2}{\rule{0.450pt}{0.400pt}}
\put(730,347.17){\rule{1.300pt}{0.400pt}}
\multiput(730.00,348.17)(3.302,-2.000){2}{\rule{0.650pt}{0.400pt}}
\multiput(736.00,345.95)(1.355,-0.447){3}{\rule{1.033pt}{0.108pt}}
\multiput(736.00,346.17)(4.855,-3.000){2}{\rule{0.517pt}{0.400pt}}
\put(743,342.17){\rule{1.300pt}{0.400pt}}
\multiput(743.00,343.17)(3.302,-2.000){2}{\rule{0.650pt}{0.400pt}}
\multiput(749.00,340.95)(1.132,-0.447){3}{\rule{0.900pt}{0.108pt}}
\multiput(749.00,341.17)(4.132,-3.000){2}{\rule{0.450pt}{0.400pt}}
\put(755,337.17){\rule{1.300pt}{0.400pt}}
\multiput(755.00,338.17)(3.302,-2.000){2}{\rule{0.650pt}{0.400pt}}
\put(761,335.17){\rule{1.500pt}{0.400pt}}
\multiput(761.00,336.17)(3.887,-2.000){2}{\rule{0.750pt}{0.400pt}}
\put(768,333.17){\rule{1.300pt}{0.400pt}}
\multiput(768.00,334.17)(3.302,-2.000){2}{\rule{0.650pt}{0.400pt}}
\put(774,331.17){\rule{1.300pt}{0.400pt}}
\multiput(774.00,332.17)(3.302,-2.000){2}{\rule{0.650pt}{0.400pt}}
\put(780,329.17){\rule{1.300pt}{0.400pt}}
\multiput(780.00,330.17)(3.302,-2.000){2}{\rule{0.650pt}{0.400pt}}
\put(786,327.17){\rule{1.300pt}{0.400pt}}
\multiput(786.00,328.17)(3.302,-2.000){2}{\rule{0.650pt}{0.400pt}}
\put(792,325.67){\rule{1.686pt}{0.400pt}}
\multiput(792.00,326.17)(3.500,-1.000){2}{\rule{0.843pt}{0.400pt}}
\put(799,324.67){\rule{1.445pt}{0.400pt}}
\multiput(799.00,325.17)(3.000,-1.000){2}{\rule{0.723pt}{0.400pt}}
\put(805,323.17){\rule{1.300pt}{0.400pt}}
\multiput(805.00,324.17)(3.302,-2.000){2}{\rule{0.650pt}{0.400pt}}
\put(811,321.67){\rule{1.445pt}{0.400pt}}
\multiput(811.00,322.17)(3.000,-1.000){2}{\rule{0.723pt}{0.400pt}}
\put(288.0,965.0){\rule[-0.200pt]{0.400pt}{0.482pt}}
\put(824,320.67){\rule{1.445pt}{0.400pt}}
\multiput(824.00,321.17)(3.000,-1.000){2}{\rule{0.723pt}{0.400pt}}
\put(817.0,322.0){\rule[-0.200pt]{1.686pt}{0.400pt}}
\put(830.0,321.0){\rule[-0.200pt]{1.445pt}{0.400pt}}
\sbox{\plotpoint}{\rule[-0.500pt]{1.000pt}{1.000pt}}%
\put(257.00,967.00){\usebox{\plotpoint}}
\put(260.77,946.61){\usebox{\plotpoint}}
\multiput(264,930)(3.607,-20.440){2}{\usebox{\plotpoint}}
\multiput(270,896)(3.713,-20.421){2}{\usebox{\plotpoint}}
\put(279.57,844.55){\usebox{\plotpoint}}
\multiput(282,832)(3.944,-20.377){2}{\usebox{\plotpoint}}
\put(292.34,783.63){\usebox{\plotpoint}}
\put(297.05,763.42){\usebox{\plotpoint}}
\multiput(301,745)(4.503,-20.261){2}{\usebox{\plotpoint}}
\put(310.68,702.66){\usebox{\plotpoint}}
\put(316.03,682.61){\usebox{\plotpoint}}
\put(321.66,662.64){\usebox{\plotpoint}}
\put(326.94,642.56){\usebox{\plotpoint}}
\put(332.42,622.55){\usebox{\plotpoint}}
\put(338.12,602.59){\usebox{\plotpoint}}
\put(344.10,582.72){\usebox{\plotpoint}}
\put(351.26,563.23){\usebox{\plotpoint}}
\put(357.82,543.54){\usebox{\plotpoint}}
\put(364.54,523.90){\usebox{\plotpoint}}
\put(372.22,504.63){\usebox{\plotpoint}}
\put(380.21,485.48){\usebox{\plotpoint}}
\multiput(382,481)(8.176,-19.077){0}{\usebox{\plotpoint}}
\put(388.29,466.36){\usebox{\plotpoint}}
\put(396.99,447.52){\usebox{\plotpoint}}
\put(406.84,429.27){\usebox{\plotpoint}}
\multiput(407,429)(9.282,-18.564){0}{\usebox{\plotpoint}}
\put(416.36,410.83){\usebox{\plotpoint}}
\multiput(419,406)(10.679,-17.798){0}{\usebox{\plotpoint}}
\put(427.06,393.06){\usebox{\plotpoint}}
\multiput(432,386)(11.513,-17.270){0}{\usebox{\plotpoint}}
\put(438.73,375.90){\usebox{\plotpoint}}
\multiput(444,368)(11.513,-17.270){0}{\usebox{\plotpoint}}
\put(450.29,358.67){\usebox{\plotpoint}}
\multiput(456,352)(13.668,-15.620){0}{\usebox{\plotpoint}}
\put(463.88,342.98){\usebox{\plotpoint}}
\multiput(469,337)(13.508,-15.759){0}{\usebox{\plotpoint}}
\put(477.59,327.41){\usebox{\plotpoint}}
\multiput(481,324)(14.676,-14.676){0}{\usebox{\plotpoint}}
\put(492.64,313.14){\usebox{\plotpoint}}
\multiput(494,312)(14.676,-14.676){0}{\usebox{\plotpoint}}
\multiput(500,306)(15.945,-13.287){0}{\usebox{\plotpoint}}
\put(508.06,299.28){\usebox{\plotpoint}}
\multiput(512,296)(16.889,-12.064){0}{\usebox{\plotpoint}}
\put(524.40,286.50){\usebox{\plotpoint}}
\multiput(525,286)(17.270,-11.513){0}{\usebox{\plotpoint}}
\multiput(531,282)(17.270,-11.513){0}{\usebox{\plotpoint}}
\put(541.82,275.25){\usebox{\plotpoint}}
\multiput(544,274)(17.270,-11.513){0}{\usebox{\plotpoint}}
\multiput(550,270)(17.270,-11.513){0}{\usebox{\plotpoint}}
\put(559.42,264.29){\usebox{\plotpoint}}
\multiput(562,263)(17.270,-11.513){0}{\usebox{\plotpoint}}
\multiput(568,259)(19.077,-8.176){0}{\usebox{\plotpoint}}
\put(577.72,254.64){\usebox{\plotpoint}}
\multiput(581,253)(17.270,-11.513){0}{\usebox{\plotpoint}}
\multiput(587,249)(18.564,-9.282){0}{\usebox{\plotpoint}}
\put(595.91,244.75){\usebox{\plotpoint}}
\multiput(600,243)(18.564,-9.282){0}{\usebox{\plotpoint}}
\multiput(606,240)(19.690,-6.563){0}{\usebox{\plotpoint}}
\put(614.93,236.53){\usebox{\plotpoint}}
\multiput(618,235)(18.564,-9.282){0}{\usebox{\plotpoint}}
\multiput(624,232)(19.077,-8.176){0}{\usebox{\plotpoint}}
\put(633.85,228.05){\usebox{\plotpoint}}
\multiput(637,227)(18.564,-9.282){0}{\usebox{\plotpoint}}
\multiput(643,224)(18.564,-9.282){0}{\usebox{\plotpoint}}
\put(652.86,219.90){\usebox{\plotpoint}}
\multiput(656,219)(18.564,-9.282){0}{\usebox{\plotpoint}}
\multiput(662,216)(19.690,-6.563){0}{\usebox{\plotpoint}}
\put(672.23,212.59){\usebox{\plotpoint}}
\multiput(674,212)(18.564,-9.282){0}{\usebox{\plotpoint}}
\multiput(680,209)(19.957,-5.702){0}{\usebox{\plotpoint}}
\put(691.65,205.45){\usebox{\plotpoint}}
\multiput(693,205)(18.564,-9.282){0}{\usebox{\plotpoint}}
\multiput(699,202)(19.690,-6.563){0}{\usebox{\plotpoint}}
\put(711.06,198.27){\usebox{\plotpoint}}
\multiput(712,198)(19.690,-6.563){0}{\usebox{\plotpoint}}
\multiput(718,196)(19.690,-6.563){0}{\usebox{\plotpoint}}
\multiput(724,194)(19.690,-6.563){0}{\usebox{\plotpoint}}
\put(730.76,191.75){\usebox{\plotpoint}}
\multiput(736,190)(19.957,-5.702){0}{\usebox{\plotpoint}}
\multiput(743,188)(20.473,-3.412){0}{\usebox{\plotpoint}}
\put(750.77,186.41){\usebox{\plotpoint}}
\multiput(755,185)(20.473,-3.412){0}{\usebox{\plotpoint}}
\multiput(761,184)(20.547,-2.935){0}{\usebox{\plotpoint}}
\put(771.10,182.48){\usebox{\plotpoint}}
\multiput(774,182)(20.473,-3.412){0}{\usebox{\plotpoint}}
\multiput(780,181)(20.473,-3.412){0}{\usebox{\plotpoint}}
\put(791.57,179.07){\usebox{\plotpoint}}
\multiput(792,179)(20.756,0.000){0}{\usebox{\plotpoint}}
\multiput(799,179)(20.756,0.000){0}{\usebox{\plotpoint}}
\multiput(805,179)(20.756,0.000){0}{\usebox{\plotpoint}}
\put(812.32,179.00){\usebox{\plotpoint}}
\multiput(817,179)(20.547,2.935){0}{\usebox{\plotpoint}}
\multiput(824,180)(20.473,3.412){0}{\usebox{\plotpoint}}
\put(832.89,181.48){\usebox{\plotpoint}}
\put(836,182){\usebox{\plotpoint}}
\sbox{\plotpoint}{\rule[-0.200pt]{0.400pt}{0.400pt}}%
\put(332,813){\usebox{\plotpoint}}
\put(332.0,574.0){\rule[-0.200pt]{0.400pt}{57.575pt}}
\put(444,677){\usebox{\plotpoint}}
\put(444.0,472.0){\rule[-0.200pt]{0.400pt}{49.384pt}}
\put(556,523){\usebox{\plotpoint}}
\put(556.0,352.0){\rule[-0.200pt]{0.400pt}{41.194pt}}
\put(668,472){\usebox{\plotpoint}}
\put(668.0,301.0){\rule[-0.200pt]{0.400pt}{41.194pt}}
\put(780,352){\usebox{\plotpoint}}
\put(780.0,215.0){\rule[-0.200pt]{0.400pt}{33.003pt}}
\put(332,694){\raisebox{-.8pt}{\makebox(0,0){$\Box$}}}
\put(444,566){\raisebox{-.8pt}{\makebox(0,0){$\Box$}}}
\put(556,438){\raisebox{-.8pt}{\makebox(0,0){$\Box$}}}
\put(668,386){\raisebox{-.8pt}{\makebox(0,0){$\Box$}}}
\put(780,287){\raisebox{-.8pt}{\makebox(0,0){$\Box$}}}
\end{picture}

%% file: fig2_inel.tex
\setlength{\unitlength}{0.240900pt}
\ifx\plotpoint\undefined\newsavebox{\plotpoint}\fi
\begin{picture}(900,809)(0,0)
\font\gnuplot=cmr10 at 10pt
\gnuplot
\sbox{\plotpoint}{\rule[-0.200pt]{0.400pt}{0.400pt}}%
\put(220.0,113.0){\rule[-0.200pt]{0.400pt}{162.126pt}}
\put(220.0,113.0){\rule[-0.200pt]{4.818pt}{0.400pt}}
\put(198,113){\makebox(0,0)[r]{0.1}}
\put(816.0,113.0){\rule[-0.200pt]{4.818pt}{0.400pt}}
\put(220.0,248.0){\rule[-0.200pt]{4.818pt}{0.400pt}}
\put(198,248){\makebox(0,0)[r]{0.15}}
\put(816.0,248.0){\rule[-0.200pt]{4.818pt}{0.400pt}}
\put(220.0,382.0){\rule[-0.200pt]{4.818pt}{0.400pt}}
\put(198,382){\makebox(0,0)[r]{0.2}}
\put(816.0,382.0){\rule[-0.200pt]{4.818pt}{0.400pt}}
\put(220.0,517.0){\rule[-0.200pt]{4.818pt}{0.400pt}}
\put(198,517){\makebox(0,0)[r]{0.25}}
\put(816.0,517.0){\rule[-0.200pt]{4.818pt}{0.400pt}}
\put(220.0,651.0){\rule[-0.200pt]{4.818pt}{0.400pt}}
\put(198,651){\makebox(0,0)[r]{0.3}}
\put(816.0,651.0){\rule[-0.200pt]{4.818pt}{0.400pt}}
\put(297.0,113.0){\rule[-0.200pt]{0.400pt}{4.818pt}}
\put(297,68){\makebox(0,0){0.5}}
\put(297.0,766.0){\rule[-0.200pt]{0.400pt}{4.818pt}}
\put(374.0,113.0){\rule[-0.200pt]{0.400pt}{4.818pt}}
\put(374,68){\makebox(0,0){1}}
\put(374.0,766.0){\rule[-0.200pt]{0.400pt}{4.818pt}}
\put(451.0,113.0){\rule[-0.200pt]{0.400pt}{4.818pt}}
\put(451,68){\makebox(0,0){1.5}}
\put(451.0,766.0){\rule[-0.200pt]{0.400pt}{4.818pt}}
\put(528.0,113.0){\rule[-0.200pt]{0.400pt}{4.818pt}}
\put(528,68){\makebox(0,0){2}}
\put(528.0,766.0){\rule[-0.200pt]{0.400pt}{4.818pt}}
\put(605.0,113.0){\rule[-0.200pt]{0.400pt}{4.818pt}}
\put(605,68){\makebox(0,0){2.5}}
\put(605.0,766.0){\rule[-0.200pt]{0.400pt}{4.818pt}}
\put(682.0,113.0){\rule[-0.200pt]{0.400pt}{4.818pt}}
\put(682,68){\makebox(0,0){3}}
\put(682.0,766.0){\rule[-0.200pt]{0.400pt}{4.818pt}}
\put(759.0,113.0){\rule[-0.200pt]{0.400pt}{4.818pt}}
\put(759,68){\makebox(0,0){3.5}}
\put(759.0,766.0){\rule[-0.200pt]{0.400pt}{4.818pt}}
\put(220.0,113.0){\rule[-0.200pt]{148.394pt}{0.400pt}}
\put(836.0,113.0){\rule[-0.200pt]{0.400pt}{162.126pt}}
\put(220.0,786.0){\rule[-0.200pt]{148.394pt}{0.400pt}}
\put(705,674){\makebox(0,0){\huge{$\epsilon$}}}
\put(528,23){\makebox(0,0){$H/\lambda$}}
\put(220.0,113.0){\rule[-0.200pt]{0.400pt}{162.126pt}}
\put(220,749){\usebox{\plotpoint}}
\put(220.00,749.00){\usebox{\plotpoint}}
\multiput(226,746)(19.690,-6.563){0}{\usebox{\plotpoint}}
\multiput(232,744)(19.077,-8.176){0}{\usebox{\plotpoint}}
\put(239.10,740.97){\usebox{\plotpoint}}
\multiput(245,739)(18.564,-9.282){0}{\usebox{\plotpoint}}
\multiput(251,736)(18.564,-9.282){0}{\usebox{\plotpoint}}
\put(258.08,732.69){\usebox{\plotpoint}}
\multiput(264,731)(18.564,-9.282){0}{\usebox{\plotpoint}}
\multiput(270,728)(19.690,-6.563){0}{\usebox{\plotpoint}}
\put(277.40,725.30){\usebox{\plotpoint}}
\multiput(282,723)(19.690,-6.563){0}{\usebox{\plotpoint}}
\multiput(288,721)(19.077,-8.176){0}{\usebox{\plotpoint}}
\put(296.49,717.25){\usebox{\plotpoint}}
\multiput(301,715)(19.690,-6.563){0}{\usebox{\plotpoint}}
\multiput(307,713)(18.564,-9.282){0}{\usebox{\plotpoint}}
\put(315.58,709.26){\usebox{\plotpoint}}
\multiput(320,708)(18.564,-9.282){0}{\usebox{\plotpoint}}
\multiput(326,705)(19.690,-6.563){0}{\usebox{\plotpoint}}
\put(334.80,701.60){\usebox{\plotpoint}}
\multiput(338,700)(19.690,-6.563){0}{\usebox{\plotpoint}}
\multiput(344,698)(19.077,-8.176){0}{\usebox{\plotpoint}}
\put(353.89,693.55){\usebox{\plotpoint}}
\multiput(357,692)(19.690,-6.563){0}{\usebox{\plotpoint}}
\multiput(363,690)(18.564,-9.282){0}{\usebox{\plotpoint}}
\put(373.09,685.83){\usebox{\plotpoint}}
\multiput(376,685)(18.564,-9.282){0}{\usebox{\plotpoint}}
\multiput(382,682)(19.690,-6.563){0}{\usebox{\plotpoint}}
\put(392.20,677.90){\usebox{\plotpoint}}
\multiput(394,677)(18.564,-9.282){0}{\usebox{\plotpoint}}
\multiput(400,674)(19.957,-5.702){0}{\usebox{\plotpoint}}
\put(411.25,669.88){\usebox{\plotpoint}}
\multiput(413,669)(19.690,-6.563){0}{\usebox{\plotpoint}}
\multiput(419,667)(18.564,-9.282){0}{\usebox{\plotpoint}}
\put(430.54,662.42){\usebox{\plotpoint}}
\multiput(432,662)(18.564,-9.282){0}{\usebox{\plotpoint}}
\multiput(438,659)(18.564,-9.282){0}{\usebox{\plotpoint}}
\put(449.53,654.16){\usebox{\plotpoint}}
\multiput(450,654)(18.564,-9.282){0}{\usebox{\plotpoint}}
\multiput(456,651)(19.957,-5.702){0}{\usebox{\plotpoint}}
\put(468.61,646.20){\usebox{\plotpoint}}
\multiput(469,646)(19.690,-6.563){0}{\usebox{\plotpoint}}
\multiput(475,644)(18.564,-9.282){0}{\usebox{\plotpoint}}
\put(487.69,638.13){\usebox{\plotpoint}}
\multiput(488,638)(19.690,-6.563){0}{\usebox{\plotpoint}}
\multiput(494,636)(18.564,-9.282){0}{\usebox{\plotpoint}}
\multiput(500,633)(19.690,-6.563){0}{\usebox{\plotpoint}}
\put(506.95,630.52){\usebox{\plotpoint}}
\multiput(512,628)(19.957,-5.702){0}{\usebox{\plotpoint}}
\multiput(519,626)(18.564,-9.282){0}{\usebox{\plotpoint}}
\put(526.07,622.64){\usebox{\plotpoint}}
\multiput(531,621)(18.564,-9.282){0}{\usebox{\plotpoint}}
\multiput(537,618)(19.077,-8.176){0}{\usebox{\plotpoint}}
\put(545.17,614.61){\usebox{\plotpoint}}
\multiput(550,613)(18.564,-9.282){0}{\usebox{\plotpoint}}
\multiput(556,610)(19.690,-6.563){0}{\usebox{\plotpoint}}
\put(564.35,606.82){\usebox{\plotpoint}}
\multiput(568,605)(19.957,-5.702){0}{\usebox{\plotpoint}}
\multiput(575,603)(18.564,-9.282){0}{\usebox{\plotpoint}}
\put(583.40,598.80){\usebox{\plotpoint}}
\multiput(587,597)(19.690,-6.563){0}{\usebox{\plotpoint}}
\multiput(593,595)(19.077,-8.176){0}{\usebox{\plotpoint}}
\put(602.65,591.12){\usebox{\plotpoint}}
\multiput(606,590)(18.564,-9.282){0}{\usebox{\plotpoint}}
\multiput(612,587)(19.690,-6.563){0}{\usebox{\plotpoint}}
\put(621.75,583.12){\usebox{\plotpoint}}
\multiput(624,582)(19.077,-8.176){0}{\usebox{\plotpoint}}
\multiput(631,579)(19.690,-6.563){0}{\usebox{\plotpoint}}
\put(640.85,575.08){\usebox{\plotpoint}}
\multiput(643,574)(19.690,-6.563){0}{\usebox{\plotpoint}}
\multiput(649,572)(19.077,-8.176){0}{\usebox{\plotpoint}}
\put(660.18,567.61){\usebox{\plotpoint}}
\multiput(662,567)(18.564,-9.282){0}{\usebox{\plotpoint}}
\multiput(668,564)(18.564,-9.282){0}{\usebox{\plotpoint}}
\put(679.14,559.29){\usebox{\plotpoint}}
\multiput(680,559)(19.077,-8.176){0}{\usebox{\plotpoint}}
\multiput(687,556)(19.690,-6.563){0}{\usebox{\plotpoint}}
\put(698.29,551.36){\usebox{\plotpoint}}
\multiput(699,551)(19.690,-6.563){0}{\usebox{\plotpoint}}
\multiput(705,549)(19.077,-8.176){0}{\usebox{\plotpoint}}
\put(717.38,543.31){\usebox{\plotpoint}}
\multiput(718,543)(19.690,-6.563){0}{\usebox{\plotpoint}}
\multiput(724,541)(18.564,-9.282){0}{\usebox{\plotpoint}}
\multiput(730,538)(19.690,-6.563){0}{\usebox{\plotpoint}}
\put(736.65,535.72){\usebox{\plotpoint}}
\multiput(743,533)(19.690,-6.563){0}{\usebox{\plotpoint}}
\multiput(749,531)(18.564,-9.282){0}{\usebox{\plotpoint}}
\put(755.77,527.74){\usebox{\plotpoint}}
\multiput(761,526)(19.077,-8.176){0}{\usebox{\plotpoint}}
\multiput(768,523)(18.564,-9.282){0}{\usebox{\plotpoint}}
\put(774.88,519.71){\usebox{\plotpoint}}
\multiput(780,518)(18.564,-9.282){0}{\usebox{\plotpoint}}
\multiput(786,515)(19.690,-6.563){0}{\usebox{\plotpoint}}
\put(794.13,512.09){\usebox{\plotpoint}}
\multiput(799,510)(19.690,-6.563){0}{\usebox{\plotpoint}}
\multiput(805,508)(18.564,-9.282){0}{\usebox{\plotpoint}}
\put(813.17,503.91){\usebox{\plotpoint}}
\multiput(817,502)(19.957,-5.702){0}{\usebox{\plotpoint}}
\multiput(824,500)(18.564,-9.282){0}{\usebox{\plotpoint}}
\put(832.36,496.21){\usebox{\plotpoint}}
\put(836,495){\usebox{\plotpoint}}
\put(220,593){\usebox{\plotpoint}}
\put(220,591.17){\rule{1.300pt}{0.400pt}}
\multiput(220.00,592.17)(3.302,-2.000){2}{\rule{0.650pt}{0.400pt}}
\multiput(226.00,589.95)(1.132,-0.447){3}{\rule{0.900pt}{0.108pt}}
\multiput(226.00,590.17)(4.132,-3.000){2}{\rule{0.450pt}{0.400pt}}
\multiput(232.00,586.95)(1.355,-0.447){3}{\rule{1.033pt}{0.108pt}}
\multiput(232.00,587.17)(4.855,-3.000){2}{\rule{0.517pt}{0.400pt}}
\put(239,583.17){\rule{1.300pt}{0.400pt}}
\multiput(239.00,584.17)(3.302,-2.000){2}{\rule{0.650pt}{0.400pt}}
\multiput(245.00,581.95)(1.132,-0.447){3}{\rule{0.900pt}{0.108pt}}
\multiput(245.00,582.17)(4.132,-3.000){2}{\rule{0.450pt}{0.400pt}}
\put(251,578.17){\rule{1.300pt}{0.400pt}}
\multiput(251.00,579.17)(3.302,-2.000){2}{\rule{0.650pt}{0.400pt}}
\multiput(257.00,576.95)(1.355,-0.447){3}{\rule{1.033pt}{0.108pt}}
\multiput(257.00,577.17)(4.855,-3.000){2}{\rule{0.517pt}{0.400pt}}
\multiput(264.00,573.95)(1.132,-0.447){3}{\rule{0.900pt}{0.108pt}}
\multiput(264.00,574.17)(4.132,-3.000){2}{\rule{0.450pt}{0.400pt}}
\put(270,570.17){\rule{1.300pt}{0.400pt}}
\multiput(270.00,571.17)(3.302,-2.000){2}{\rule{0.650pt}{0.400pt}}
\multiput(276.00,568.95)(1.132,-0.447){3}{\rule{0.900pt}{0.108pt}}
\multiput(276.00,569.17)(4.132,-3.000){2}{\rule{0.450pt}{0.400pt}}
\put(282,565.17){\rule{1.300pt}{0.400pt}}
\multiput(282.00,566.17)(3.302,-2.000){2}{\rule{0.650pt}{0.400pt}}
\multiput(288.00,563.95)(1.355,-0.447){3}{\rule{1.033pt}{0.108pt}}
\multiput(288.00,564.17)(4.855,-3.000){2}{\rule{0.517pt}{0.400pt}}
\multiput(295.00,560.95)(1.132,-0.447){3}{\rule{0.900pt}{0.108pt}}
\multiput(295.00,561.17)(4.132,-3.000){2}{\rule{0.450pt}{0.400pt}}
\put(301,557.17){\rule{1.300pt}{0.400pt}}
\multiput(301.00,558.17)(3.302,-2.000){2}{\rule{0.650pt}{0.400pt}}
\multiput(307.00,555.95)(1.132,-0.447){3}{\rule{0.900pt}{0.108pt}}
\multiput(307.00,556.17)(4.132,-3.000){2}{\rule{0.450pt}{0.400pt}}
\multiput(313.00,552.95)(1.355,-0.447){3}{\rule{1.033pt}{0.108pt}}
\multiput(313.00,553.17)(4.855,-3.000){2}{\rule{0.517pt}{0.400pt}}
\put(320,549.17){\rule{1.300pt}{0.400pt}}
\multiput(320.00,550.17)(3.302,-2.000){2}{\rule{0.650pt}{0.400pt}}
\multiput(326.00,547.95)(1.132,-0.447){3}{\rule{0.900pt}{0.108pt}}
\multiput(326.00,548.17)(4.132,-3.000){2}{\rule{0.450pt}{0.400pt}}
\put(332,544.17){\rule{1.300pt}{0.400pt}}
\multiput(332.00,545.17)(3.302,-2.000){2}{\rule{0.650pt}{0.400pt}}
\multiput(338.00,542.95)(1.132,-0.447){3}{\rule{0.900pt}{0.108pt}}
\multiput(338.00,543.17)(4.132,-3.000){2}{\rule{0.450pt}{0.400pt}}
\multiput(344.00,539.95)(1.355,-0.447){3}{\rule{1.033pt}{0.108pt}}
\multiput(344.00,540.17)(4.855,-3.000){2}{\rule{0.517pt}{0.400pt}}
\put(351,536.17){\rule{1.300pt}{0.400pt}}
\multiput(351.00,537.17)(3.302,-2.000){2}{\rule{0.650pt}{0.400pt}}
\multiput(357.00,534.95)(1.132,-0.447){3}{\rule{0.900pt}{0.108pt}}
\multiput(357.00,535.17)(4.132,-3.000){2}{\rule{0.450pt}{0.400pt}}
\put(363,531.17){\rule{1.300pt}{0.400pt}}
\multiput(363.00,532.17)(3.302,-2.000){2}{\rule{0.650pt}{0.400pt}}
\multiput(369.00,529.95)(1.355,-0.447){3}{\rule{1.033pt}{0.108pt}}
\multiput(369.00,530.17)(4.855,-3.000){2}{\rule{0.517pt}{0.400pt}}
\multiput(376.00,526.95)(1.132,-0.447){3}{\rule{0.900pt}{0.108pt}}
\multiput(376.00,527.17)(4.132,-3.000){2}{\rule{0.450pt}{0.400pt}}
\put(382,523.17){\rule{1.300pt}{0.400pt}}
\multiput(382.00,524.17)(3.302,-2.000){2}{\rule{0.650pt}{0.400pt}}
\multiput(388.00,521.95)(1.132,-0.447){3}{\rule{0.900pt}{0.108pt}}
\multiput(388.00,522.17)(4.132,-3.000){2}{\rule{0.450pt}{0.400pt}}
\put(394,518.17){\rule{1.300pt}{0.400pt}}
\multiput(394.00,519.17)(3.302,-2.000){2}{\rule{0.650pt}{0.400pt}}
\multiput(400.00,516.95)(1.355,-0.447){3}{\rule{1.033pt}{0.108pt}}
\multiput(400.00,517.17)(4.855,-3.000){2}{\rule{0.517pt}{0.400pt}}
\multiput(407.00,513.95)(1.132,-0.447){3}{\rule{0.900pt}{0.108pt}}
\multiput(407.00,514.17)(4.132,-3.000){2}{\rule{0.450pt}{0.400pt}}
\put(413,510.17){\rule{1.300pt}{0.400pt}}
\multiput(413.00,511.17)(3.302,-2.000){2}{\rule{0.650pt}{0.400pt}}
\multiput(419.00,508.95)(1.132,-0.447){3}{\rule{0.900pt}{0.108pt}}
\multiput(419.00,509.17)(4.132,-3.000){2}{\rule{0.450pt}{0.400pt}}
\multiput(425.00,505.95)(1.355,-0.447){3}{\rule{1.033pt}{0.108pt}}
\multiput(425.00,506.17)(4.855,-3.000){2}{\rule{0.517pt}{0.400pt}}
\put(432,502.17){\rule{1.300pt}{0.400pt}}
\multiput(432.00,503.17)(3.302,-2.000){2}{\rule{0.650pt}{0.400pt}}
\multiput(438.00,500.95)(1.132,-0.447){3}{\rule{0.900pt}{0.108pt}}
\multiput(438.00,501.17)(4.132,-3.000){2}{\rule{0.450pt}{0.400pt}}
\put(444,497.17){\rule{1.300pt}{0.400pt}}
\multiput(444.00,498.17)(3.302,-2.000){2}{\rule{0.650pt}{0.400pt}}
\multiput(450.00,495.95)(1.132,-0.447){3}{\rule{0.900pt}{0.108pt}}
\multiput(450.00,496.17)(4.132,-3.000){2}{\rule{0.450pt}{0.400pt}}
\multiput(456.00,492.95)(1.355,-0.447){3}{\rule{1.033pt}{0.108pt}}
\multiput(456.00,493.17)(4.855,-3.000){2}{\rule{0.517pt}{0.400pt}}
\put(463,489.17){\rule{1.300pt}{0.400pt}}
\multiput(463.00,490.17)(3.302,-2.000){2}{\rule{0.650pt}{0.400pt}}
\multiput(469.00,487.95)(1.132,-0.447){3}{\rule{0.900pt}{0.108pt}}
\multiput(469.00,488.17)(4.132,-3.000){2}{\rule{0.450pt}{0.400pt}}
\put(475,484.17){\rule{1.300pt}{0.400pt}}
\multiput(475.00,485.17)(3.302,-2.000){2}{\rule{0.650pt}{0.400pt}}
\multiput(481.00,482.95)(1.355,-0.447){3}{\rule{1.033pt}{0.108pt}}
\multiput(481.00,483.17)(4.855,-3.000){2}{\rule{0.517pt}{0.400pt}}
\multiput(488.00,479.95)(1.132,-0.447){3}{\rule{0.900pt}{0.108pt}}
\multiput(488.00,480.17)(4.132,-3.000){2}{\rule{0.450pt}{0.400pt}}
\put(494,476.17){\rule{1.300pt}{0.400pt}}
\multiput(494.00,477.17)(3.302,-2.000){2}{\rule{0.650pt}{0.400pt}}
\multiput(500.00,474.95)(1.132,-0.447){3}{\rule{0.900pt}{0.108pt}}
\multiput(500.00,475.17)(4.132,-3.000){2}{\rule{0.450pt}{0.400pt}}
\put(506,471.17){\rule{1.300pt}{0.400pt}}
\multiput(506.00,472.17)(3.302,-2.000){2}{\rule{0.650pt}{0.400pt}}
\multiput(512.00,469.95)(1.355,-0.447){3}{\rule{1.033pt}{0.108pt}}
\multiput(512.00,470.17)(4.855,-3.000){2}{\rule{0.517pt}{0.400pt}}
\multiput(519.00,466.95)(1.132,-0.447){3}{\rule{0.900pt}{0.108pt}}
\multiput(519.00,467.17)(4.132,-3.000){2}{\rule{0.450pt}{0.400pt}}
\put(525,463.17){\rule{1.300pt}{0.400pt}}
\multiput(525.00,464.17)(3.302,-2.000){2}{\rule{0.650pt}{0.400pt}}
\multiput(531.00,461.95)(1.132,-0.447){3}{\rule{0.900pt}{0.108pt}}
\multiput(531.00,462.17)(4.132,-3.000){2}{\rule{0.450pt}{0.400pt}}
\put(537,458.17){\rule{1.500pt}{0.400pt}}
\multiput(537.00,459.17)(3.887,-2.000){2}{\rule{0.750pt}{0.400pt}}
\multiput(544.00,456.95)(1.132,-0.447){3}{\rule{0.900pt}{0.108pt}}
\multiput(544.00,457.17)(4.132,-3.000){2}{\rule{0.450pt}{0.400pt}}
\multiput(550.00,453.95)(1.132,-0.447){3}{\rule{0.900pt}{0.108pt}}
\multiput(550.00,454.17)(4.132,-3.000){2}{\rule{0.450pt}{0.400pt}}
\put(556,450.17){\rule{1.300pt}{0.400pt}}
\multiput(556.00,451.17)(3.302,-2.000){2}{\rule{0.650pt}{0.400pt}}
\multiput(562.00,448.95)(1.132,-0.447){3}{\rule{0.900pt}{0.108pt}}
\multiput(562.00,449.17)(4.132,-3.000){2}{\rule{0.450pt}{0.400pt}}
\multiput(568.00,445.95)(1.355,-0.447){3}{\rule{1.033pt}{0.108pt}}
\multiput(568.00,446.17)(4.855,-3.000){2}{\rule{0.517pt}{0.400pt}}
\put(575,442.17){\rule{1.300pt}{0.400pt}}
\multiput(575.00,443.17)(3.302,-2.000){2}{\rule{0.650pt}{0.400pt}}
\multiput(581.00,440.95)(1.132,-0.447){3}{\rule{0.900pt}{0.108pt}}
\multiput(581.00,441.17)(4.132,-3.000){2}{\rule{0.450pt}{0.400pt}}
\put(587,437.17){\rule{1.300pt}{0.400pt}}
\multiput(587.00,438.17)(3.302,-2.000){2}{\rule{0.650pt}{0.400pt}}
\multiput(593.00,435.95)(1.355,-0.447){3}{\rule{1.033pt}{0.108pt}}
\multiput(593.00,436.17)(4.855,-3.000){2}{\rule{0.517pt}{0.400pt}}
\multiput(600.00,432.95)(1.132,-0.447){3}{\rule{0.900pt}{0.108pt}}
\multiput(600.00,433.17)(4.132,-3.000){2}{\rule{0.450pt}{0.400pt}}
\put(606,429.17){\rule{1.300pt}{0.400pt}}
\multiput(606.00,430.17)(3.302,-2.000){2}{\rule{0.650pt}{0.400pt}}
\multiput(612.00,427.95)(1.132,-0.447){3}{\rule{0.900pt}{0.108pt}}
\multiput(612.00,428.17)(4.132,-3.000){2}{\rule{0.450pt}{0.400pt}}
\put(618,424.17){\rule{1.300pt}{0.400pt}}
\multiput(618.00,425.17)(3.302,-2.000){2}{\rule{0.650pt}{0.400pt}}
\multiput(624.00,422.95)(1.355,-0.447){3}{\rule{1.033pt}{0.108pt}}
\multiput(624.00,423.17)(4.855,-3.000){2}{\rule{0.517pt}{0.400pt}}
\multiput(631.00,419.95)(1.132,-0.447){3}{\rule{0.900pt}{0.108pt}}
\multiput(631.00,420.17)(4.132,-3.000){2}{\rule{0.450pt}{0.400pt}}
\put(637,416.17){\rule{1.300pt}{0.400pt}}
\multiput(637.00,417.17)(3.302,-2.000){2}{\rule{0.650pt}{0.400pt}}
\multiput(643.00,414.95)(1.132,-0.447){3}{\rule{0.900pt}{0.108pt}}
\multiput(643.00,415.17)(4.132,-3.000){2}{\rule{0.450pt}{0.400pt}}
\put(649,411.17){\rule{1.500pt}{0.400pt}}
\multiput(649.00,412.17)(3.887,-2.000){2}{\rule{0.750pt}{0.400pt}}
\multiput(656.00,409.95)(1.132,-0.447){3}{\rule{0.900pt}{0.108pt}}
\multiput(656.00,410.17)(4.132,-3.000){2}{\rule{0.450pt}{0.400pt}}
\multiput(662.00,406.95)(1.132,-0.447){3}{\rule{0.900pt}{0.108pt}}
\multiput(662.00,407.17)(4.132,-3.000){2}{\rule{0.450pt}{0.400pt}}
\put(668,403.17){\rule{1.300pt}{0.400pt}}
\multiput(668.00,404.17)(3.302,-2.000){2}{\rule{0.650pt}{0.400pt}}
\multiput(674.00,401.95)(1.132,-0.447){3}{\rule{0.900pt}{0.108pt}}
\multiput(674.00,402.17)(4.132,-3.000){2}{\rule{0.450pt}{0.400pt}}
\multiput(680.00,398.95)(1.355,-0.447){3}{\rule{1.033pt}{0.108pt}}
\multiput(680.00,399.17)(4.855,-3.000){2}{\rule{0.517pt}{0.400pt}}
\put(687,395.17){\rule{1.300pt}{0.400pt}}
\multiput(687.00,396.17)(3.302,-2.000){2}{\rule{0.650pt}{0.400pt}}
\multiput(693.00,393.95)(1.132,-0.447){3}{\rule{0.900pt}{0.108pt}}
\multiput(693.00,394.17)(4.132,-3.000){2}{\rule{0.450pt}{0.400pt}}
\put(699,390.17){\rule{1.300pt}{0.400pt}}
\multiput(699.00,391.17)(3.302,-2.000){2}{\rule{0.650pt}{0.400pt}}
\multiput(705.00,388.95)(1.355,-0.447){3}{\rule{1.033pt}{0.108pt}}
\multiput(705.00,389.17)(4.855,-3.000){2}{\rule{0.517pt}{0.400pt}}
\multiput(712.00,385.95)(1.132,-0.447){3}{\rule{0.900pt}{0.108pt}}
\multiput(712.00,386.17)(4.132,-3.000){2}{\rule{0.450pt}{0.400pt}}
\put(718,382.17){\rule{1.300pt}{0.400pt}}
\multiput(718.00,383.17)(3.302,-2.000){2}{\rule{0.650pt}{0.400pt}}
\multiput(724.00,380.95)(1.132,-0.447){3}{\rule{0.900pt}{0.108pt}}
\multiput(724.00,381.17)(4.132,-3.000){2}{\rule{0.450pt}{0.400pt}}
\put(730,377.17){\rule{1.300pt}{0.400pt}}
\multiput(730.00,378.17)(3.302,-2.000){2}{\rule{0.650pt}{0.400pt}}
\multiput(736.00,375.95)(1.355,-0.447){3}{\rule{1.033pt}{0.108pt}}
\multiput(736.00,376.17)(4.855,-3.000){2}{\rule{0.517pt}{0.400pt}}
\multiput(743.00,372.95)(1.132,-0.447){3}{\rule{0.900pt}{0.108pt}}
\multiput(743.00,373.17)(4.132,-3.000){2}{\rule{0.450pt}{0.400pt}}
\put(749,369.17){\rule{1.300pt}{0.400pt}}
\multiput(749.00,370.17)(3.302,-2.000){2}{\rule{0.650pt}{0.400pt}}
\multiput(755.00,367.95)(1.132,-0.447){3}{\rule{0.900pt}{0.108pt}}
\multiput(755.00,368.17)(4.132,-3.000){2}{\rule{0.450pt}{0.400pt}}
\put(761,364.17){\rule{1.500pt}{0.400pt}}
\multiput(761.00,365.17)(3.887,-2.000){2}{\rule{0.750pt}{0.400pt}}
\multiput(768.00,362.95)(1.132,-0.447){3}{\rule{0.900pt}{0.108pt}}
\multiput(768.00,363.17)(4.132,-3.000){2}{\rule{0.450pt}{0.400pt}}
\multiput(774.00,359.95)(1.132,-0.447){3}{\rule{0.900pt}{0.108pt}}
\multiput(774.00,360.17)(4.132,-3.000){2}{\rule{0.450pt}{0.400pt}}
\put(780,356.17){\rule{1.300pt}{0.400pt}}
\multiput(780.00,357.17)(3.302,-2.000){2}{\rule{0.650pt}{0.400pt}}
\multiput(786.00,354.95)(1.132,-0.447){3}{\rule{0.900pt}{0.108pt}}
\multiput(786.00,355.17)(4.132,-3.000){2}{\rule{0.450pt}{0.400pt}}
\multiput(792.00,351.95)(1.355,-0.447){3}{\rule{1.033pt}{0.108pt}}
\multiput(792.00,352.17)(4.855,-3.000){2}{\rule{0.517pt}{0.400pt}}
\put(799,348.17){\rule{1.300pt}{0.400pt}}
\multiput(799.00,349.17)(3.302,-2.000){2}{\rule{0.650pt}{0.400pt}}
\multiput(805.00,346.95)(1.132,-0.447){3}{\rule{0.900pt}{0.108pt}}
\multiput(805.00,347.17)(4.132,-3.000){2}{\rule{0.450pt}{0.400pt}}
\put(811,343.17){\rule{1.300pt}{0.400pt}}
\multiput(811.00,344.17)(3.302,-2.000){2}{\rule{0.650pt}{0.400pt}}
\multiput(817.00,341.95)(1.355,-0.447){3}{\rule{1.033pt}{0.108pt}}
\multiput(817.00,342.17)(4.855,-3.000){2}{\rule{0.517pt}{0.400pt}}
\multiput(824.00,338.95)(1.132,-0.447){3}{\rule{0.900pt}{0.108pt}}
\multiput(824.00,339.17)(4.132,-3.000){2}{\rule{0.450pt}{0.400pt}}
\put(830,335.17){\rule{1.300pt}{0.400pt}}
\multiput(830.00,336.17)(3.302,-2.000){2}{\rule{0.650pt}{0.400pt}}
\sbox{\plotpoint}{\rule[-0.500pt]{1.000pt}{1.000pt}}%
\put(220,372){\usebox{\plotpoint}}
\put(220.00,372.00){\usebox{\plotpoint}}
\multiput(226,369)(19.690,-6.563){0}{\usebox{\plotpoint}}
\multiput(232,367)(19.077,-8.176){0}{\usebox{\plotpoint}}
\put(239.10,363.97){\usebox{\plotpoint}}
\multiput(245,362)(18.564,-9.282){0}{\usebox{\plotpoint}}
\multiput(251,359)(19.690,-6.563){0}{\usebox{\plotpoint}}
\put(258.38,356.41){\usebox{\plotpoint}}
\multiput(264,354)(18.564,-9.282){0}{\usebox{\plotpoint}}
\multiput(270,351)(19.690,-6.563){0}{\usebox{\plotpoint}}
\put(277.44,348.28){\usebox{\plotpoint}}
\multiput(282,346)(19.690,-6.563){0}{\usebox{\plotpoint}}
\multiput(288,344)(19.077,-8.176){0}{\usebox{\plotpoint}}
\put(296.63,340.46){\usebox{\plotpoint}}
\multiput(301,339)(18.564,-9.282){0}{\usebox{\plotpoint}}
\multiput(307,336)(18.564,-9.282){0}{\usebox{\plotpoint}}
\put(315.63,332.25){\usebox{\plotpoint}}
\multiput(320,331)(18.564,-9.282){0}{\usebox{\plotpoint}}
\multiput(326,328)(19.690,-6.563){0}{\usebox{\plotpoint}}
\put(334.84,324.58){\usebox{\plotpoint}}
\multiput(338,323)(19.690,-6.563){0}{\usebox{\plotpoint}}
\multiput(344,321)(19.077,-8.176){0}{\usebox{\plotpoint}}
\put(354.11,316.96){\usebox{\plotpoint}}
\multiput(357,316)(18.564,-9.282){0}{\usebox{\plotpoint}}
\multiput(363,313)(18.564,-9.282){0}{\usebox{\plotpoint}}
\put(373.13,308.82){\usebox{\plotpoint}}
\multiput(376,308)(18.564,-9.282){0}{\usebox{\plotpoint}}
\multiput(382,305)(19.690,-6.563){0}{\usebox{\plotpoint}}
\put(392.24,300.88){\usebox{\plotpoint}}
\multiput(394,300)(19.690,-6.563){0}{\usebox{\plotpoint}}
\multiput(400,298)(19.077,-8.176){0}{\usebox{\plotpoint}}
\put(411.34,292.83){\usebox{\plotpoint}}
\multiput(413,292)(19.690,-6.563){0}{\usebox{\plotpoint}}
\multiput(419,290)(18.564,-9.282){0}{\usebox{\plotpoint}}
\put(430.64,285.39){\usebox{\plotpoint}}
\multiput(432,285)(18.564,-9.282){0}{\usebox{\plotpoint}}
\multiput(438,282)(19.690,-6.563){0}{\usebox{\plotpoint}}
\put(449.64,277.18){\usebox{\plotpoint}}
\multiput(450,277)(18.564,-9.282){0}{\usebox{\plotpoint}}
\multiput(456,274)(19.957,-5.702){0}{\usebox{\plotpoint}}
\put(468.69,269.15){\usebox{\plotpoint}}
\multiput(469,269)(19.690,-6.563){0}{\usebox{\plotpoint}}
\multiput(475,267)(18.564,-9.282){0}{\usebox{\plotpoint}}
\multiput(481,264)(19.957,-5.702){0}{\usebox{\plotpoint}}
\put(488.09,261.96){\usebox{\plotpoint}}
\multiput(494,259)(18.564,-9.282){0}{\usebox{\plotpoint}}
\multiput(500,256)(19.690,-6.563){0}{\usebox{\plotpoint}}
\put(507.00,253.50){\usebox{\plotpoint}}
\multiput(512,251)(19.957,-5.702){0}{\usebox{\plotpoint}}
\multiput(519,249)(18.564,-9.282){0}{\usebox{\plotpoint}}
\put(526.11,245.63){\usebox{\plotpoint}}
\multiput(531,244)(18.564,-9.282){0}{\usebox{\plotpoint}}
\multiput(537,241)(19.077,-8.176){0}{\usebox{\plotpoint}}
\put(545.21,237.60){\usebox{\plotpoint}}
\multiput(550,236)(18.564,-9.282){0}{\usebox{\plotpoint}}
\multiput(556,233)(19.690,-6.563){0}{\usebox{\plotpoint}}
\put(564.39,229.80){\usebox{\plotpoint}}
\multiput(568,228)(19.957,-5.702){0}{\usebox{\plotpoint}}
\multiput(575,226)(18.564,-9.282){0}{\usebox{\plotpoint}}
\put(583.60,222.13){\usebox{\plotpoint}}
\multiput(587,221)(18.564,-9.282){0}{\usebox{\plotpoint}}
\multiput(593,218)(19.077,-8.176){0}{\usebox{\plotpoint}}
\put(602.70,214.10){\usebox{\plotpoint}}
\multiput(606,213)(18.564,-9.282){0}{\usebox{\plotpoint}}
\multiput(612,210)(19.690,-6.563){0}{\usebox{\plotpoint}}
\put(621.79,206.10){\usebox{\plotpoint}}
\multiput(624,205)(19.957,-5.702){0}{\usebox{\plotpoint}}
\multiput(631,203)(18.564,-9.282){0}{\usebox{\plotpoint}}
\put(640.85,198.08){\usebox{\plotpoint}}
\multiput(643,197)(19.690,-6.563){0}{\usebox{\plotpoint}}
\multiput(649,195)(19.077,-8.176){0}{\usebox{\plotpoint}}
\put(660.18,190.61){\usebox{\plotpoint}}
\multiput(662,190)(18.564,-9.282){0}{\usebox{\plotpoint}}
\multiput(668,187)(19.690,-6.563){0}{\usebox{\plotpoint}}
\put(679.19,182.40){\usebox{\plotpoint}}
\multiput(680,182)(19.077,-8.176){0}{\usebox{\plotpoint}}
\multiput(687,179)(19.690,-6.563){0}{\usebox{\plotpoint}}
\put(698.29,174.36){\usebox{\plotpoint}}
\multiput(699,174)(19.690,-6.563){0}{\usebox{\plotpoint}}
\multiput(705,172)(19.077,-8.176){0}{\usebox{\plotpoint}}
\put(717.71,167.10){\usebox{\plotpoint}}
\multiput(718,167)(18.564,-9.282){0}{\usebox{\plotpoint}}
\multiput(724,164)(18.564,-9.282){0}{\usebox{\plotpoint}}
\multiput(730,161)(19.690,-6.563){0}{\usebox{\plotpoint}}
\put(736.65,158.72){\usebox{\plotpoint}}
\multiput(743,156)(19.690,-6.563){0}{\usebox{\plotpoint}}
\multiput(749,154)(18.564,-9.282){0}{\usebox{\plotpoint}}
\put(755.77,150.74){\usebox{\plotpoint}}
\multiput(761,149)(19.077,-8.176){0}{\usebox{\plotpoint}}
\multiput(768,146)(19.690,-6.563){0}{\usebox{\plotpoint}}
\put(775.17,143.42){\usebox{\plotpoint}}
\multiput(780,141)(18.564,-9.282){0}{\usebox{\plotpoint}}
\multiput(786,138)(19.690,-6.563){0}{\usebox{\plotpoint}}
\put(794.13,135.09){\usebox{\plotpoint}}
\multiput(799,133)(19.690,-6.563){0}{\usebox{\plotpoint}}
\multiput(805,131)(18.564,-9.282){0}{\usebox{\plotpoint}}
\put(813.30,127.23){\usebox{\plotpoint}}
\multiput(817,126)(19.077,-8.176){0}{\usebox{\plotpoint}}
\multiput(824,123)(18.564,-9.282){0}{\usebox{\plotpoint}}
\put(832.40,119.20){\usebox{\plotpoint}}
\put(836,118){\usebox{\plotpoint}}
\sbox{\plotpoint}{\rule[-0.200pt]{0.400pt}{0.400pt}}%
\put(389,665){\usebox{\plotpoint}}
\put(389.0,301.0){\rule[-0.200pt]{0.400pt}{87.688pt}}
\put(389,490){\raisebox{-.8pt}{\makebox(0,0){$\Box$}}}
\end{picture}